\documentclass{tien-shan2006}
\usepackage{times}
\usepackage{epsfig}
\begin{document}
\title[Status and recent results of the Pierre Auger Observatory] %
{Status and recent results of the Pierre Auger Observatory}

\author[Pierre Auger Collaboration]%
{Serguei Vorobiov for the Pierre Auger Collaboration}
\presenter{Laboratoire AstroParticule et Cosmologie \& IN2P3/CNRS, Coll\`ege de France,
11, place Marcelin Berthelot, 75231 Paris C\'edex 05, France.
E-mail: vorobiov@in2p3.fr}

\maketitle

\begin{abstract}
The Pierre Auger Observatory aims to determine the nature and origin of the ultra-high energy cosmic rays (UHECR). The Auger hybrid detector combines fluorescence observations of extended air showers, initiated in the atmosphere by these most energetic particles, with measures of the shower front at the ground level by its large array of Cherenkov water tanks. This allows to improve considerably the precision on reconstructed primary cosmic ray parameters, and to make important cross-calibrations between two techniques at these energies, unreachable with accelerator experiments. The Southern Auger site in Argentina is close to completion. The first measure of the primary cosmic rays energy spectrum, the anisotropy search results, and the limit on the photon fraction in the UHECR are discussed.
\end{abstract}

\section[Introduction]{Objectives of the ultra-high energy cosmic ray physics}

Despite the recent considerable progress in cosmic ray physics domain~\cite{Hillas0607109}, the key questions concerning ultra-high energy ($\geq 10^{17}~\rm{eV}$) cosmic rays are still waiting for an answer. The observation in the cosmic rays of particles with energies reaching~$10^{20}~\rm{eV}$ limits the list of potential astrophysical acceleration sites to the most energetic objects in the Universe like active galactic nuclei, gamma-ray bursts or galaxy clusters. There are also more exotic scenarios, in which these highest energy particles result from interactions or decay of the primordial Universe relics such as topological defects or super-heavy dark matter. 

The first question to clarify is whether the nearly isotropic cosmic ray flux is suppressed at energies above~$10^{19.8}~\rm{eV}$, as one would expect in the standard acceleration scenario with the nuclei arriving from sources at cosmological distances. This spectral feature (the GZK cut-off) corresponds to the effective threshold of pion production in the interaction of the UHECR protons with the CMB radiation. At similar energies, nuclei photo-dissociate on the CMB. As a consequence, the horizon of the UHECR sources in the standard scenario is restricted to our local ``neighbourhood'' ($\simeq$ 50 Mpc). Analysis of arrival directions of events and study of anisotropies at small and large angular scales may further help to distinguish between different types of sources, and to provide constraints on extragalactic magnetic field strength and the UHECR primary particle charge. 

The composition studies are of crucial importance for the test of the UHECR production models. For example, limits on the photon fraction at the highest energies may constrain strongly exotic UHECR models. Measurements of shower properties in the atmosphere (depth of the maximum of cascade) and at the ground (thickness and curvature of the shower front, muon richness etc.) are confronted with simulations to provide necessary discrimination criteria. Another important issue of these studies is to understand the systematic difference in energy measurement between previous results obtained using different experimental techniques: from the particle density at the ground (AGASA), or from the fluorescence (HiRes) or Cherenkov (Yakutsk) emission triggered by UHECR-induced showers in the atmosphere.

The Pierre Auger Observatory was designed~\cite{Auger,AugerDesignReport} to answer these key questions of the UHECR physics. The project aims at large aperture ($> 7000~\rm{km}^2 \rm{sr}$ above $10^{19}~\rm{eV}$) hybrid detection (combining air fluorescence and ground particle techniques) of the highest energy cosmic rays with the full-sky exposure (with 1 site per hemisphere). After presenting the Southern Auger site and its current status, we will describe the first Auger results on the UHECR spectrum, anisotropy studies, and the limit on the photon contents in the cosmic rays.

\section[Auger]{The Pierre Auger Observatory}

\begin{figure}[]
\begin{minipage}[]{0.55\textwidth}
\begin{center}
\epsfig{figure=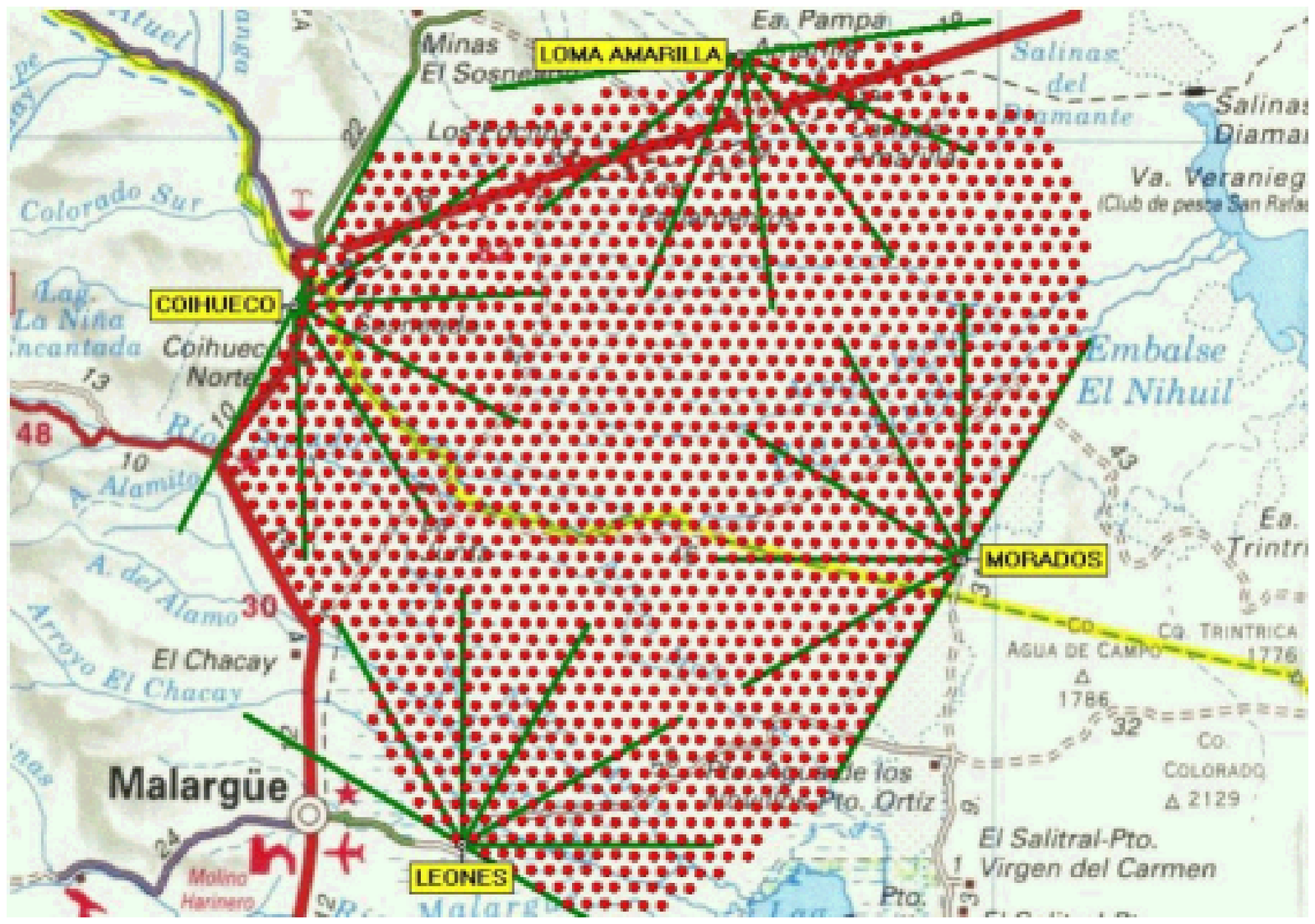,width=0.95\textwidth}
\end{center}
\end{minipage}
\begin{minipage}[]{0.45\textwidth}
\begin{center}
\epsfig{figure=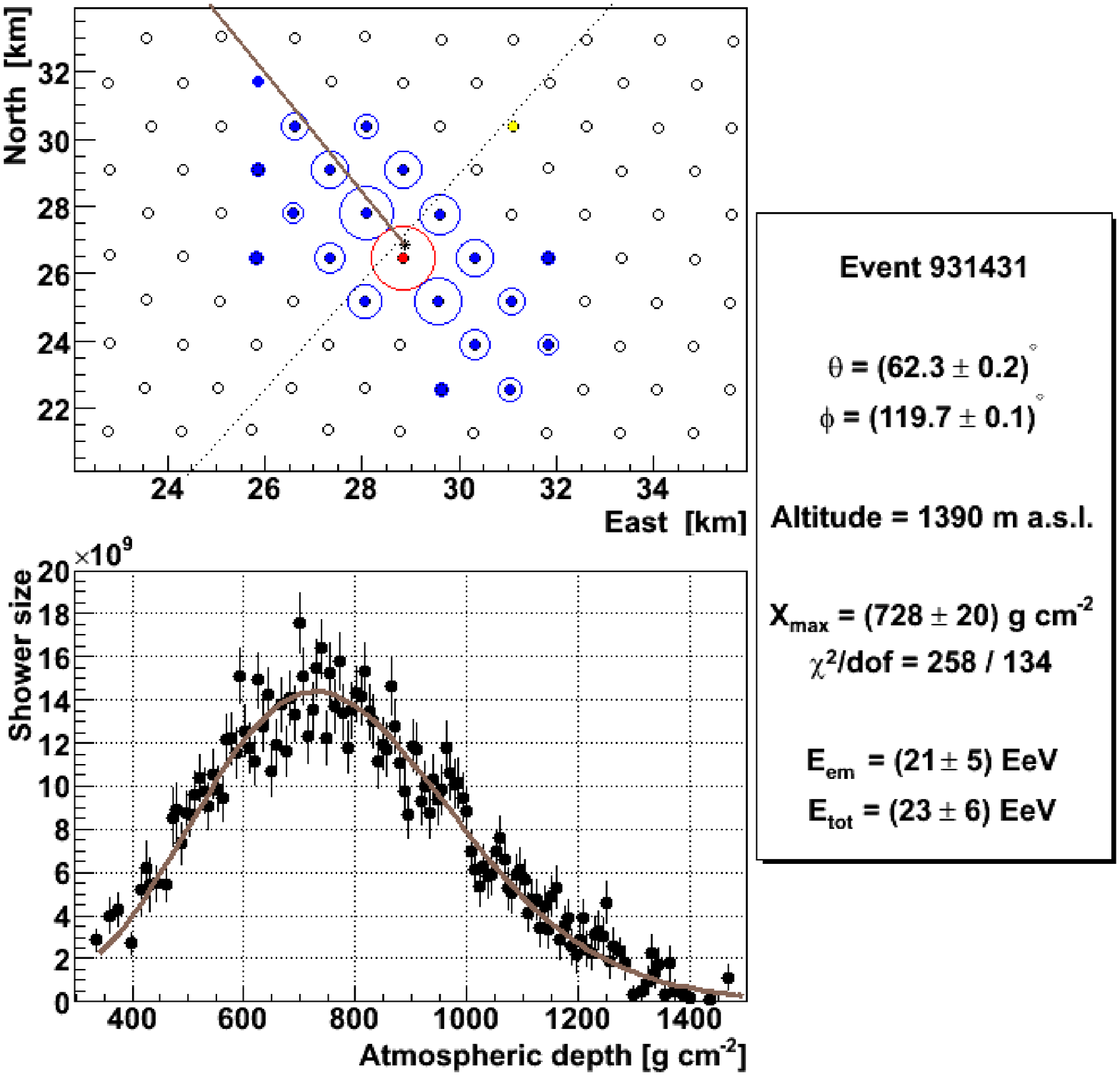,width=1.0\textwidth}
\end{center}
\end{minipage}
\caption{\it  (left) \rm The plan of the Auger Observatory, with 1600 Surface Detector stations and 4 Fluorescence Detector sites. \it (right)\rm  ~High energy hybrid event. In the upper plot, each triggered station is marked by a blue circle. The bold line is the projection of the shower axis on the ground, with the core near the station with maximal signal. The fluorescence detector captured the side view of the shower (the dotted line is the intersection of the Shower-Detector Plane with the horizontal plane containing the eye). The lower plot shows the reconstructed shower longitudinal profile.}
\label{fig:AugerHybridDetector}
\end{figure}

The Southern Auger site is located near Malarg\"ue, in the Mendoza region of Argentina. Its hybrid design (see the figure~\ref{fig:AugerHybridDetector}) allows the simultaneous detection of the same cosmic ray events by two complementary techniques. After its completion, the Auger Surface Detector (SD) will consist of 1600 12-tons water Cherenkov tanks spaced in 1.5~$\rm{km}$ triangular array on the $3000~\rm{km}^2$ area. Three photomuliplier tubes (PMTs) in each tank collect the Cherenkov light from the passage of the electromagnetic and muonic components of showers through the purified water. The PMTs signals are digitized at~$40~\rm{MHz}$ sampling frequency, which provides a temporal resolution of~$8~\rm{ns}$. Signal timing information and the integrated charge values are used for reconstruction of shower geometry (arrival direction and core position). Then an energy estimate can be obtained by comparing the lateral distribution of signal with simulations.

The Auger Fluorescence Detector (FD), once completed, will be composed of 4 fluorescence sites located on the edges of the surface array, with 6 Schmidt design telescopes per site, covering each a~$30^\circ$ range in azimuth (see the figure~\ref{fig:AugerHybridDetector}) and~$0^\circ - 30^\circ$ range in elevation. Each telescope consists of~$11~\rm{m}^2$ segmented spherical mirrors (radius of curvature~3.4~$\rm{m}$), focusing the light from the~$2.2~\rm{m}$ diameter diaphragm onto a camera of~$20\times22$ PMTs. The image of a shower developing in the field of view of a telescope represents a track of triggered PMTs, which enables to reconstruct a shower-detector plane with a high precision ($\simeq 0.3^\circ$). When, in addition to an FD telescope, one or more SD tanks participate in the event (hybrid detection), the SD timing information improves considerably the shower geometry reconstruction. Then, with the help of the absolute calibration of camera pixels and the knowledge of shower geometry, one can estimate the primary energy from the total amount of fluorescence light. This estimate is nearly calorimetric as it is related directly to ionization loss by electrons and positrons in the showers, and only a small correction ($\simeq 10\%$ at Auger energies) should take into account the ``missing'' energy due to muons and neutrinos. An even more valuable information concerning energy determination is obtained from the so-called ``Golden'' hybrid events, like that one shown in the figure~\ref{fig:AugerHybridDetector}, when the same event can be reconstructed independently by the SD and the FD.  

Currently, nearly 1000 ($\simeq 60\%$) SD stations are fully operational, and about 200 more tanks are deployed in the pampa. 18 out of 24 FD telescopes (3 sites) are now completed, and are taking data during clear moonless nights. More details on the Detector performances and calibration, numerous facilities for atmospheric monitoring can be found in the Proceedings of the~$29^{th}$ ICRC at Pune, India (2005) (see~\cite{AugerAtICRC2005} for the compilation). 

\begin{figure}[]
\begin{minipage}[]{0.45\textwidth}
\begin{center}
\epsfig{figure=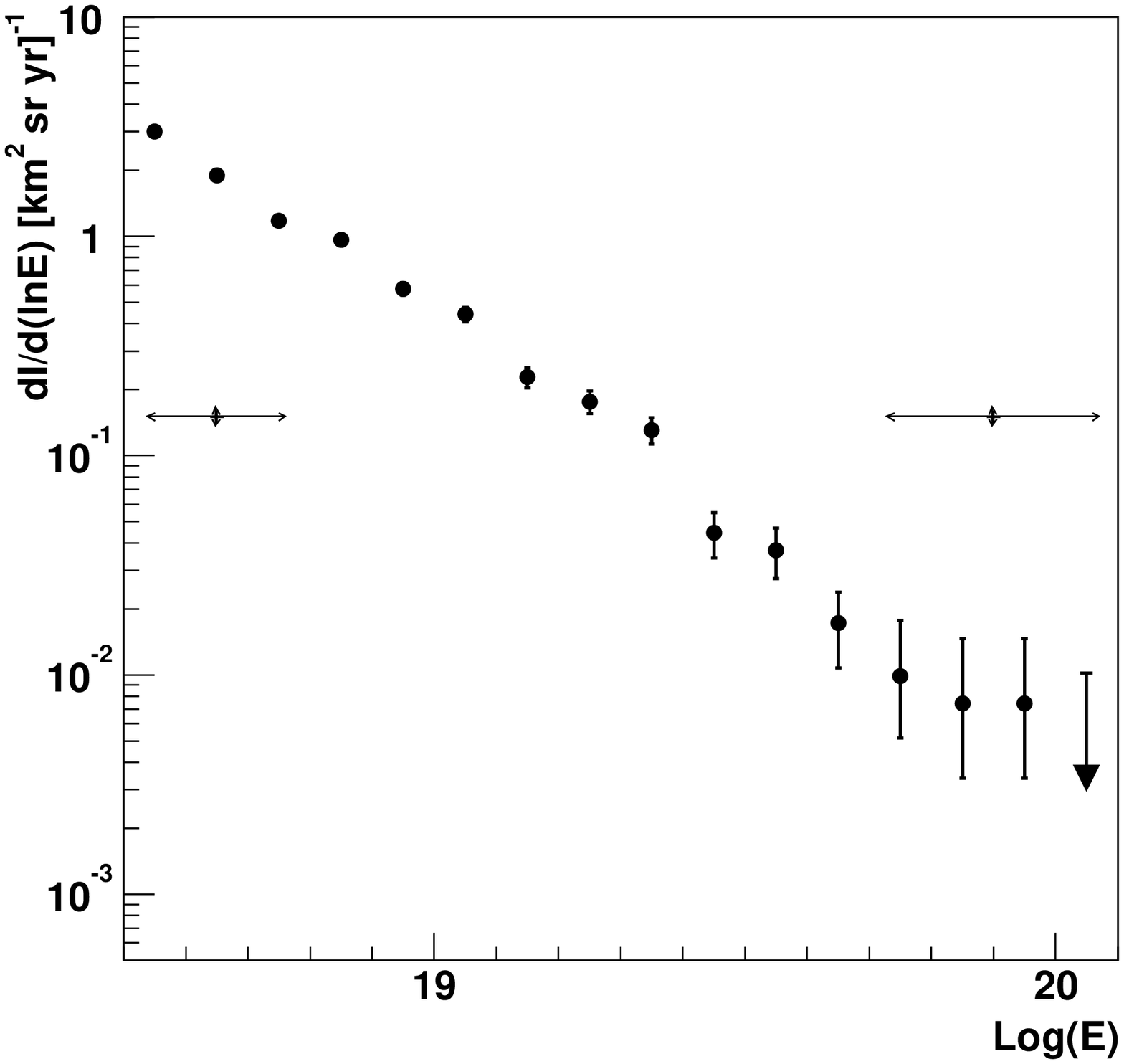,width=0.95\textwidth}
\end{center}
\end{minipage}
\begin{minipage}[]{0.55\textwidth}
\begin{center}
\epsfig{figure=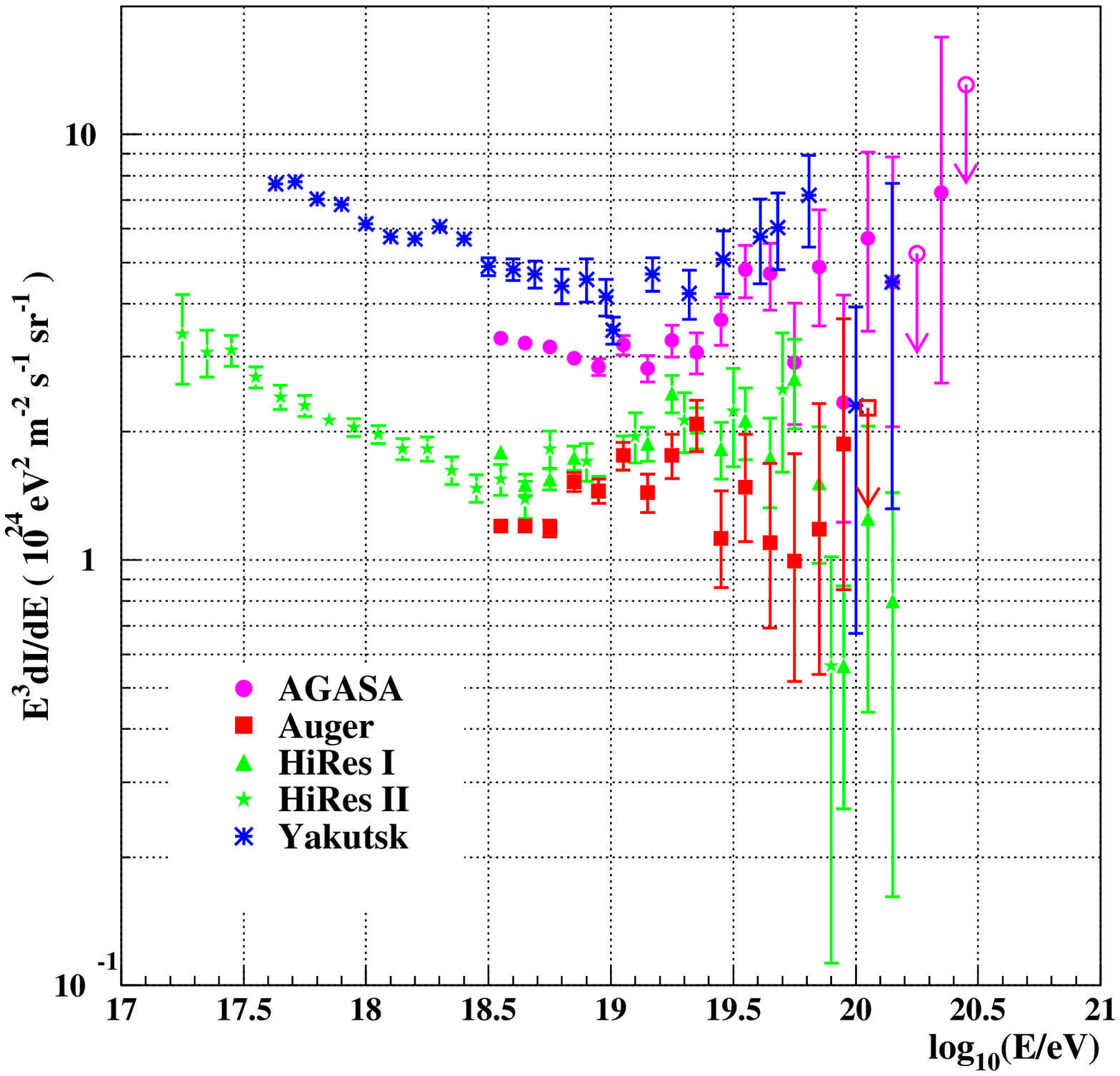,width=1.\textwidth}
\end{center}
\end{minipage}
\caption{\it  (left) \rm Energy spectrum by Auger (in~$E\frac{dI}{dE}$ representation). Errors on data points indicate statistical uncertainty (or 95\% CL upper limit). Two error bars indicate the systematic uncertainty at two different energies. \it (right) \rm ~The measured Auger spectrum (in~$E^3\frac{dI}{dE}$ representation) is superimposed on the results of the previous experiments~\cite{AGASASpectrum,HiResSpectrum,YakutskSpectrum}.}
\label{fig:UHECRSpectrum}
\end{figure}

\section[Spectrum]{Measurement of the primary cosmic ray energy spectrum above 3 EeV}

The data taken from 1 January 2004 to 5 June 2005 were used to make a first estimate~\cite{AugerSpectrumICRC2005} of the primary UHECR spectrum. While the average Auger SD array size for this period was only~$\sim{22}\%$ of the planned~$3000~\rm{km}^2$, the integrated exposure was already similar to those by the largest previous experiments. The method used to derive the spectrum is almost free of any assumptions on primary UHECR composition or hadronic interaction models. It allows to combine the power of large SD array aperture with the nearly calorimetric FD energy measurement.

The energy assignment to the SD events is a two-step procedure. First, we establish for each event an energy-related parameter. We use for this purpose S(1000), the estimate of integrated signal size at a distance of 1000~$m$ from the shower axis, which is determined from a fit of the lateral distribution of signal of all tanks triggered by the shower. Simulations show that S(1000) is almost proportional to primary energy and that at these distances from the axis the fluctuations of lateral distribution are minimal for studied energies and adopted array geometry. The constant intensity cut method, which exploits the nearly isotropy of cosmic rays, is used to rescale S(1000) value from different shower inclinations. At the second step, a rule for converting S(1000) to energy is established using a subset of high quality ``Golden'' hybrid events.

To derive the spectrum, we used 3525 events above 3~EeV, detected at zenith angles~$\theta < 60^\circ$ and falling within the well-defined fiducial area. The simulations (reinforced by an independent check of trigger probability with the hybrid events) show that the array acceptance for such showers is~100\%. Consequently, the array exposure for the selected dataset is simply defined by the array geometry and the live-time of SD tanks. The spectrum, obtained by dividing the number of events in energy bins by the exposure value ($1750~\rm{km}^2~\rm{sr}~\rm{yr}$), is shown on the left plot of figure~\ref{fig:UHECRSpectrum}.

The large part of indicated systematic errors in energy assignment comes from the limited number of hybrid events that were used to establish the S(1000) -- energy conversion, especially at the highest energies. This systematics will of course shrink rapidly with the growing amount of data. Another large sources of systematics from the FD energy scale itself ($\sim{25\%}$ in total) are the uncertainty in fluorescence yield ($15\%$) and the absolute calibration of the FD telescopes ($12\%$). Both cited uncertainty values will also be reduced in the near future. However, at this early stage of experiment, one cannot conclude yet on the presence of the GZK cutoff in the spectrum. 

The measured spectrum is shown on the right plot of figure~\ref{fig:UHECRSpectrum} with the results of previous experiments. The Auger data points are~$\sim{10\%}$ below the HiRes flux measures. It should be also noted that preliminary studies based on SD event simulations provide energies that are systematically (by~$\sim 25\%$) higher than those derived from the FD calibration. The Pierre Auger Observatory, with its large statistics and the rich information available for each shower, will investigate this intriguing difference. 

\section[Anisotropy]{Anisotropy Studies}

\begin{figure}[]
\begin{minipage}[]{0.6\textwidth}
\begin{center}
\epsfig{width=1.\textwidth,file=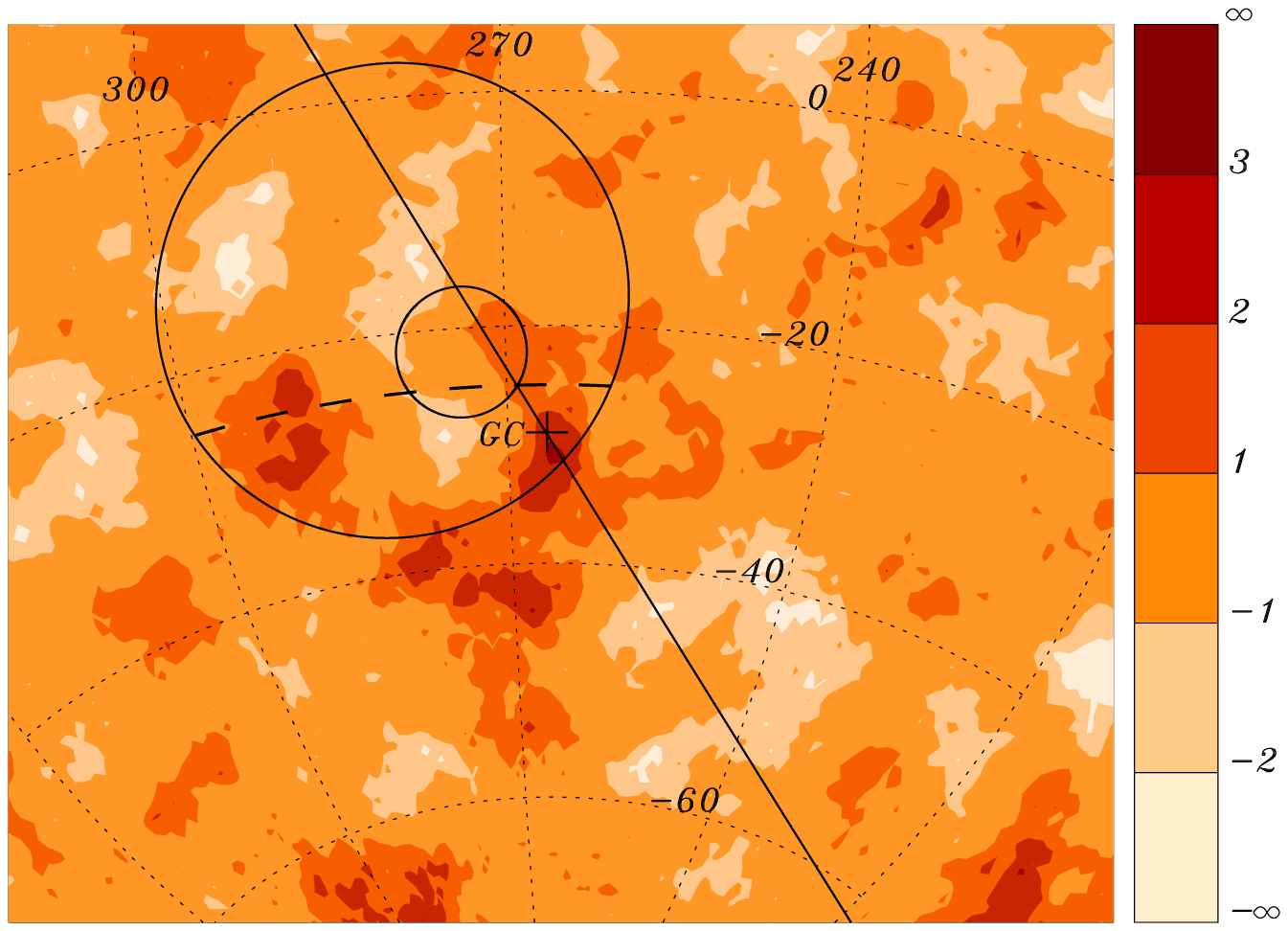}
\end{center}
\end{minipage}
\begin{minipage}[]{0.4\textwidth}
\begin{center}
\epsfig{width=1.\textwidth,file=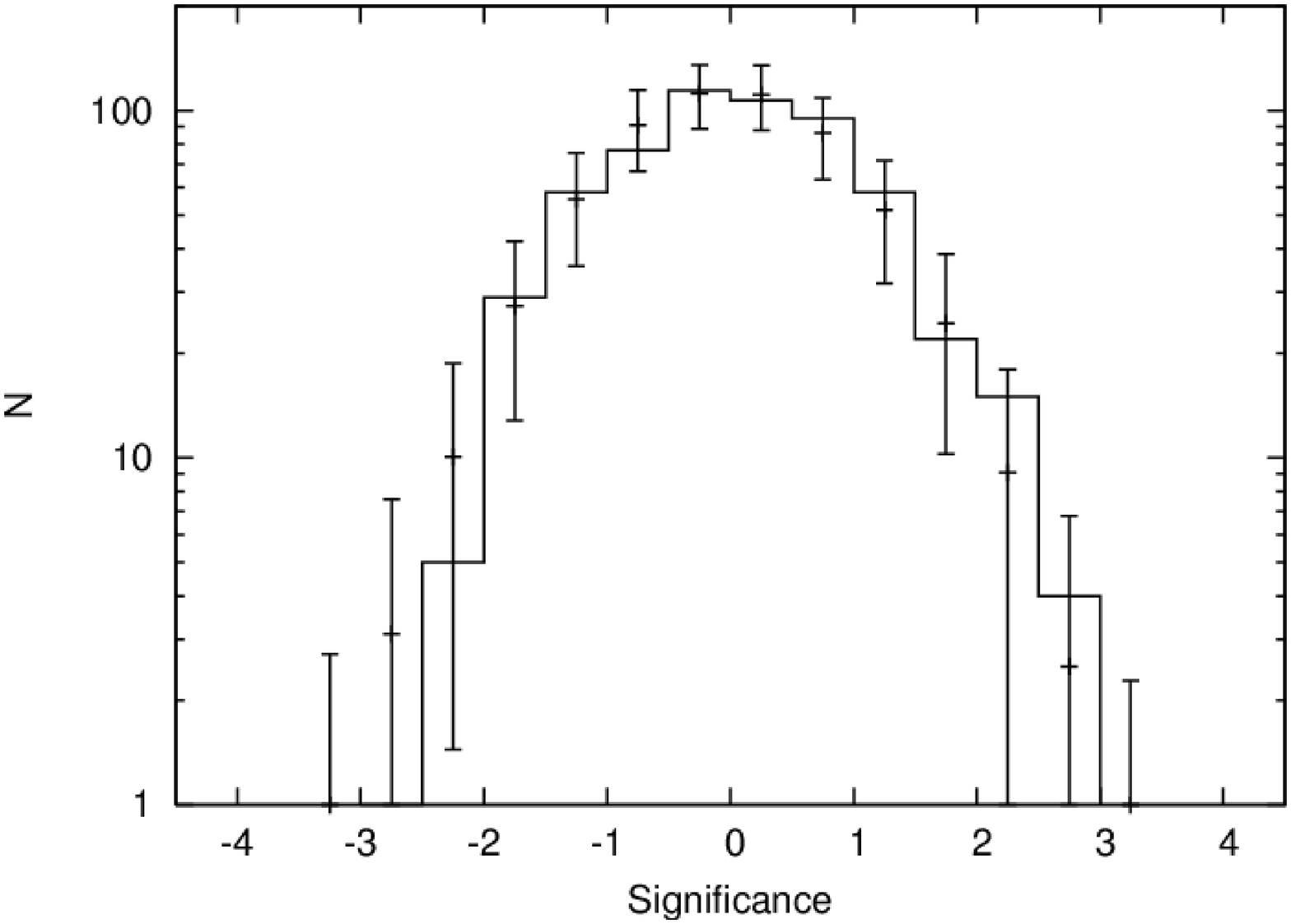}
\end{center}
\end{minipage}
\caption{\it{(left)} \rm The Auger map of CR overdensity significances near the GC region (top-hat
  windows of $5^\circ$  radius). The GC position is marked by a cross on the galactic plane. The large circle represents the AGASA excess region, the dashed line being their field of view limit, and the small circle -- the SUGAR excess region. \it{(right)} \rm The corresponding histogram of overdensities computed on a grid of
$3^\circ$ spacing compared to isotropic expectations (points with $2\sigma$ bounds).}
\label{fig:GCResults}
\end{figure}

The data from 1 January 2004 to 30 March 2006 were analyzed~\cite{AugerGC} to search for anisotropies near the Galactic Center (GC). This region represents an attractive target for such studies, as it harbors the very massive black hole, and the detection of a very close TeV~$\gamma$-ray source by H.E.S.S. collaboration~\cite{hess} had given birth to a number of theoretical models predicting the flux of neutrons at EeV energies. In addition, recent H.E.S.S. detection of the diffuse~$\gamma$-ray flux, correlated with the giant molecular clouds near the GC~\cite{hess2}, has also provided the indirect evidence for the cosmic ray acceleration in that region, though at much lower energies. There were previous claims by the AGASA and SUGAR collaborations of significant event excesses at EeV energies near to the GC, for [1--2.5]~EeV and [0.8--3.2]~EeV energy bands, respectively. The GC passes at~$\sim{6^\circ}$ from the zenith at the Auger South site latitude, and our dataset for the GC studies (79265 SD and 3439 hybrid events with similar energies $10^{17.9}\ 
{\rm eV}<E<10^{18.5}$~eV) is significantly larger than that of AGASA or SUGAR. 

The figure~\ref{fig:GCResults} summarizes the results of our anisotropy analysis for the Galactic Center region. The relative exposure of the different sky directions (coverage map) has been cross-checked using two different techniques leading to a difference of~$\sim{0.5\%}$ in background estimate, a level well below the Poissonian fluctuations and the excesses to test. Then the significance of eventual anisotropies in the UHECR arrival direction distribution was estimated by comparison of the observed number of events with that expected from an isotropic cosmic ray flux. The significance maps were built in circular windows of $5^\circ$ radius. This angular scale is convenient to visualize the overdensity distributions in the windows studied by SUGAR (excess size $\sim{5^\circ}$) and AGASA (excess size $\sim{20^\circ}$). Additional tests have been made with modified energy ranges to take into account possible differences in energy calibration. In all cases, no significant excess has been found in Auger data. We therefore do not confirm the excesses observed by AGASA and SUGAR. In addition, we have set a limit on a point-like source at the GC using the datasets of SD-only and hybrid events (the latter yielding excellent resolution of~$\sim{0.6^\circ}$).  

A scan for correlations of cosmic ray arrival directions with the galactic plane and super-galactic plane has been made, but with a smaller data set, at energies in the [1-5]~EeV range and above 5~EeV, and no significant excess has been found~\cite{AugerGCICRC2005}. A blind search for overdensities in the cosmic ray flux for the same energy ranges and at two angular scales of $5^\circ$ and $15^\circ$ has also given the results consistent with isotropy~\cite{AugerBRICRC2005}.

\section[Photon Limit]{Upper limit to the photon fraction in cosmic rays above~10~EeV}

\begin{figure}[]
\begin{minipage}[]{0.5\textwidth}
\begin{center}
\epsfig{figure=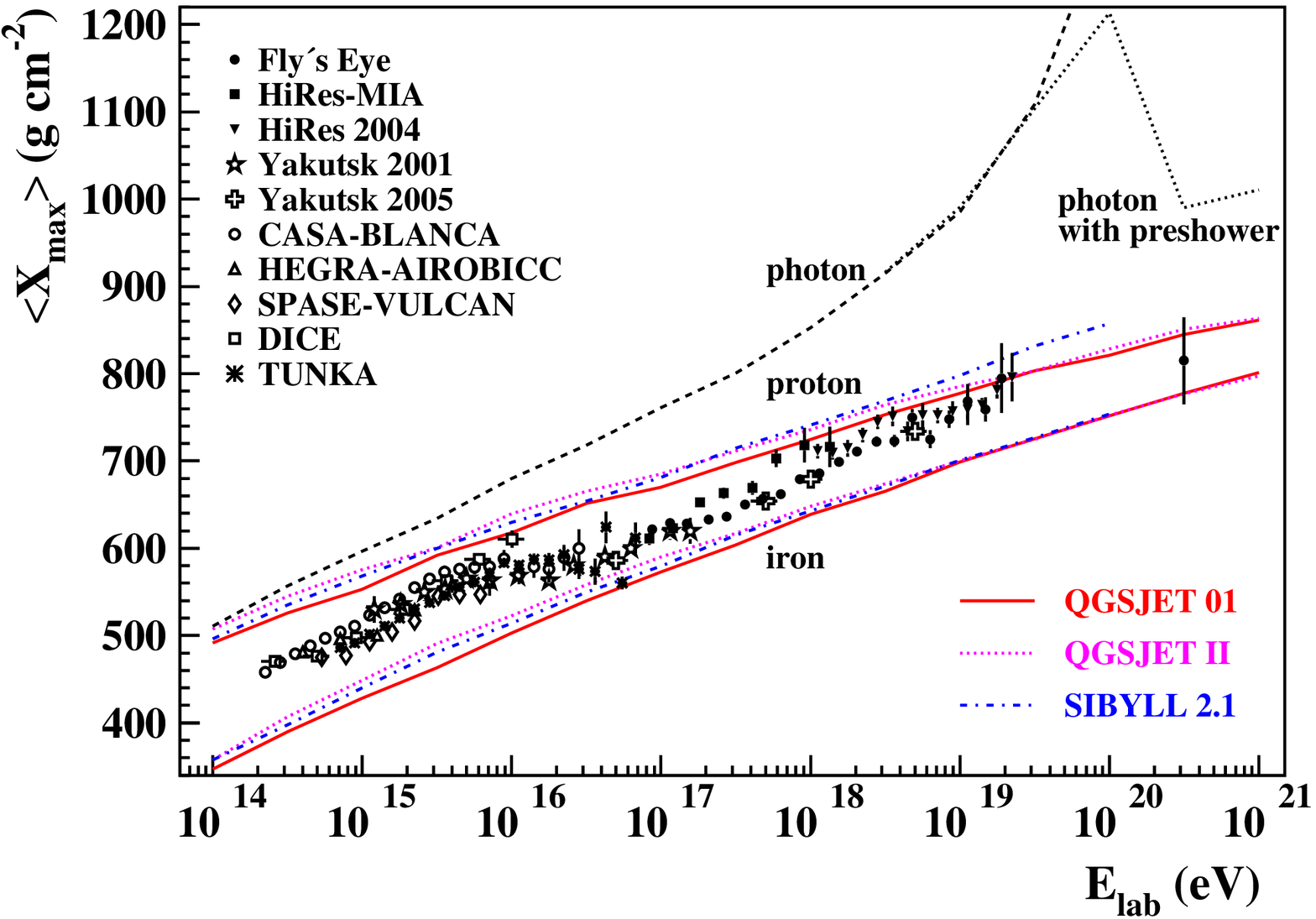,width=1.\textwidth}
\end{center}
\end{minipage}
\begin{minipage}[]{0.5\textwidth}
\begin{center}
\epsfig{figure=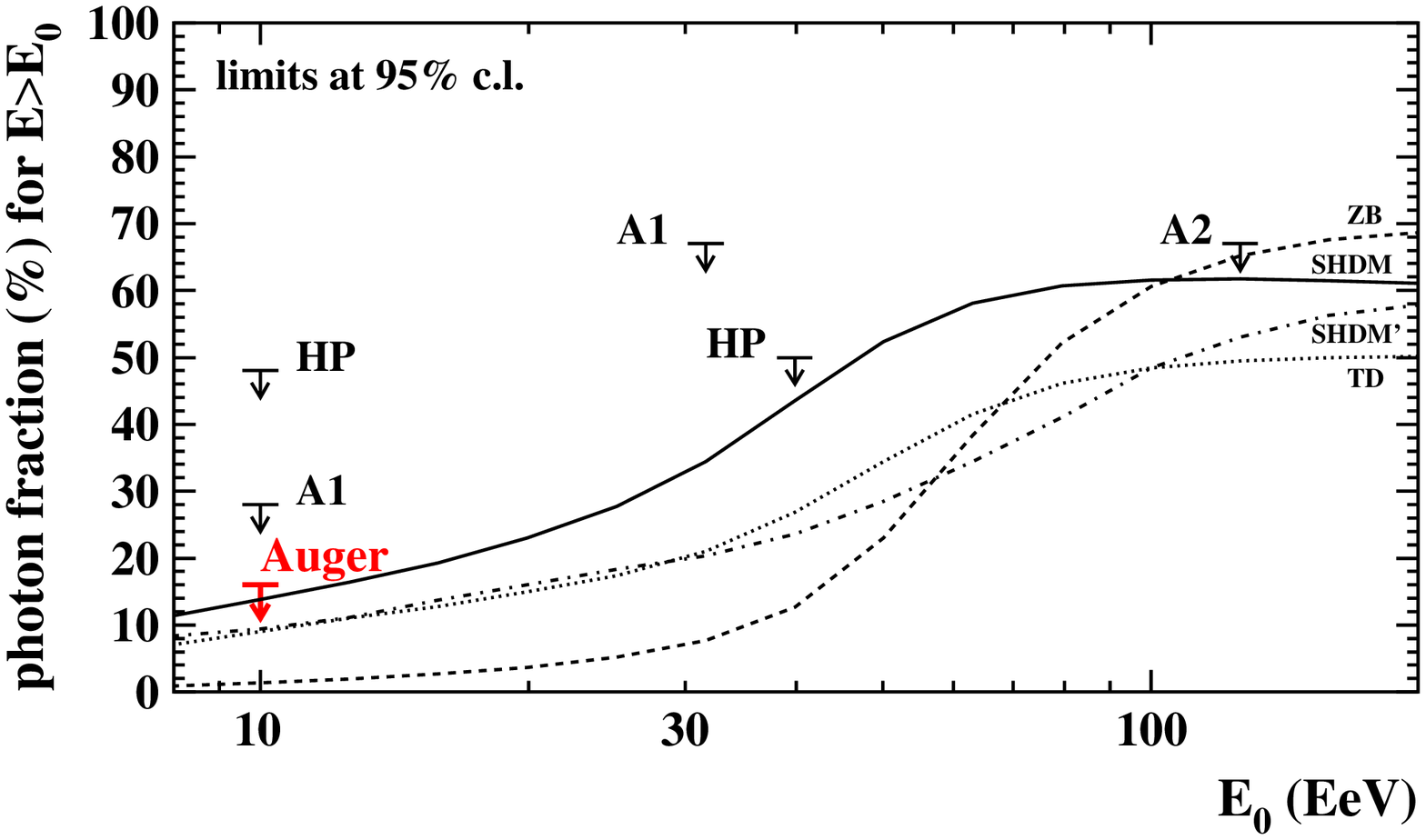,width=1.\textwidth}
\end{center}
\end{minipage}
\caption{\it  (left) \rm Average depth of shower maximum $X_{max}$ as a function of energy for simulated photons, protons and iron nuclei. The experimental data points are also shown. \it (right)\rm  ~The resulting upper limit on the photon fraction, together with the previous results~\cite{HPPhotonLimit,AGASAPhotonLimit,AGASAPhotonLimitRisse05} obtained from Haverah Park and AGASA data. The curves show predictions of exotic models, calculated on the assumption of AGASA spectrum. See~\cite{AugerPhotonLimit} and the references therein for more details.}
\label{fig:PhotonLimitStaff}
\end{figure}

The Auger hybrid data from January 2004 to February 2006 were used to set a limit to the photon fraction in the UHECR above 10~EeV~\cite{AugerPhotonLimit}. For the first time, such a limit is obtained from direct measurement of the depth of shower maximum,~$X_{max}$. At these energies, the development of photon-induced showers is delayed by the LPM effect. Consequently, the average~$X_{max}$ values for photon showers are much larger (by~$\simeq{200}~\rm{g/cm^2}$) than those for hadron-induced showers (see the figure~\ref{fig:PhotonLimitStaff}). The fluctuations of~$X_{max}$ values for photon showers are also large ($\simeq{80}~\rm{g/cm^2}$ at~10~EeV). At even higher energies ($\geq{50}$~EeV for the Southern Auger site), a competing process of photon conversion into pairs in the geomagnetic field (the ``preshowering'') brings the~$X_{max}$ values of triggered showers and the corresponding fluctuation values closer to those for hadrons. 

The average vertical depth of the atmosphere for the Southern Auger site is~$\simeq{880}~\rm{g/cm^2}$, so the nearly vertical photon showers reach the ground before their maximum development. Another consequence of the deeper shower maximum for photon primaries is the stronger attenuation of their fluorescence light for the same geometry conditions, due to the larger amount of the traversed air mass. To avoid a possible bias against photons in the detector acceptance, we have included in the analysis only sufficiently inclined events with limited maximal distance between FD telescope and shower impact point. These requirements, in addition to general hybrid quality cuts (good geometry and profile reconstruction, shower maximum observed, no clouds), assure that the acceptance ratio between photons and hadrons is not less than~80\%. The energy of the events has been calculated under assumption of their photon origin, i.e. the missing energy correction was only~$\simeq{1\%}$ due to the smallness of the photo-nuclear production cross-section. 

29 hybrid events with energies above 10~EeV satisfy the selection criteria (one of them is the ``Golden'' hybrid event shown on the figure~\ref{fig:AugerHybridDetector}). 100 photon showers were simulated for each of the real events, using the reconstructed event parameters as the input. The average $X_{max}$ values for simulated events are considerably (in the $[2\sigma ; 3.8\sigma]$ range) larger than those for the selected events. This allowed us to derive an upper limit~$f_\gamma^{ul}$ on the photon fraction. 

The calculation of the limit is based on the statistical method taking into account uncertainties $\Delta X_{max}$ in the determination of the depth of shower maximum. The main contributions to $\Delta X_{max}$ come from the uncertainties in the event geometry, longitudinal profile reconstruction and atmospheric absorption. There are also photon-specific systematic uncertainties in the $X_{max}$ value, resulting from the uncertainty in the extrapolation of the photo-nuclear production cross section, or from the uncertainty in the reconstructed energy (non-negligible due to the larger elongation rate for photons, see the figure~\ref{fig:PhotonLimitStaff}). However, the shower fluctuations for photons are considerably larger when compared to our conservative estimates of the statistical ($\simeq{28~\rm{g/cm^2}}$) and systematic ($\simeq{23~\rm{g/cm^2}}$) uncertainties in the $X_{max}$ measurement. As a consequence, the~$f_\gamma^{ul}$ value is limited mainly by the event statistics. 

For the selected data sample, the upper limit to the photon fraction is 16\% at the 95\% confidence level. The Auger limit, shown on the figure~\ref{fig:PhotonLimitStaff}, confirms and improves the previous limits obtained by the ground arrays. In the future, the larger datasets will allow to reduce this upper limit value, and to constrain strongly non-acceleration UHECR models with new upper limits at higher energies. An order of magnitude larger Surface Detector statistics will also be exploited for the photon limit studies. 

\section[Conclusions]{Conclusions and future prospects}

The first results from the Pierre Auger Observatory, though obtained with the incomplete detector over the first two years of operation out of 20 years planned, have already brought important new insights on the UHECR physics and demonstrated the power of Auger large aperture and hybrid design. There are many advances in the understanding of the detector response. To access to the large panorama of the Auger activities and results, the reader is invited to consult the 29th ICRC Proceedings~\cite{AugerAtICRC2005} and the more recent publications.

While the Southern Auger site is still under construction, a preparation work for the Northern site with even larger aperture in Colorado is on-going. At the same time, possible enhancements of the Southern site, that will allow to bring the energy threshold of the detector down to~$10^{17}$~eV and to measure with higher precision the UHECR composition in the region where the transition from galactic to extragalactic component occurs, are discussed. Several options for such enhancements are envisaged, like higher elevation angle FD telescopes, muon counters, and additional surface detectors spaced more closely (in-fill array). There are also planned R \& D on shower radio detection. We expect therefore with a reasonable optimism that many of the key questions that have been mentioned above will be answered in the near future.


\begin{thebibliography}{99}
\bibitem{Hillas0607109} A.M.~Hillas, astro-ph/0607109.
\bibitem{Auger} J.~Abraham et al., Pierre~Auger Collaboration, Nucl.~Instrum.~Meth.~A 523, 50 (2004).
\bibitem{AugerDesignReport} The Pierre Auger Project Design Report, \sf{www.auger.org/admin/DesignReport}\rm.
\bibitem{AugerAtICRC2005} \sf{www.auger.org/reports/icrc\_2005.html}\rm.
\bibitem{AugerSpectrumICRC2005} Pierre Auger Collaboration, Proc. of the 29th ICRC, Pune, India (2005) V7, p.387, [astro-ph/0507150].
\bibitem{AGASASpectrum}Takeda M. et al., Astroparticle Physics 19 (2003) 447-462.
\bibitem{HiResSpectrum} Abbasi R.U. et al. (HiRes Collaboration), Phys. Lett. B619 (2005) 271.
\bibitem{YakutskSpectrum} Egorova V.P. et al., Nucl. Phys. B (Proc. Suppl.) 136 (2004) 3-11. See also the updated web site of the Yakutsk Collaboration \sf{http://eas.ysn.ru/}\rm ~for more details.
\bibitem{AugerGC} Pierre Auger Collaboration, in press, [astro-ph/0607382].
\bibitem{hess} F.~Aharonian et al. (H.E.S.S. Collaboration), Astron. Astrophys. 425 (2004) L13, [astro-ph/0408145].
\bibitem{hess2} F.~Aharonian et al. (H.E.S.S. Collaboration), Nature 439 (2006) 695, [astro-ph/0603021].
\bibitem{AugerGCICRC2005} Pierre Auger Collaboration, Proc. of the 29th ICRC, Pune, India (2005) V7, p. 67, [astro-ph/0507331].
\bibitem{AugerBRICRC2005} Pierre Auger Collaboration, Proc. of the 29th ICRC, Pune, India (2005) V7, p. 75, [astro-ph/0507600].
\bibitem{AugerPhotonLimit} Pierre Auger Collaboration, submitted to Astroparticle Physics, [astro-ph/0606619].
\bibitem{HPPhotonLimit}M.~Ave et al., Phys.~Rev.~Lett.~85, 2244 (2000); Phys.~Rev.~D65, 063007 (2002).
\bibitem{AGASAPhotonLimit} K.~Shinozaki et al., Astrophys.~J.~571, L117 (2002).
\bibitem{AGASAPhotonLimitRisse05} M.~Risse et al., Phys.~Rev.~Lett. 95, 171102 (2005).
\end{thebibliography}
\end{document}